\documentclass[twocolumn,showpacs,pra,amsmath,amssymb,
 aps,superscriptaddress]{revtex4-1}
\usepackage{graphicx}

\newcommand{\ket}[1]{|{#1}\rangle}			
\newcommand{\bra}[1]{\langle{#1}|}			

\begin{document}
\title{Robustness of quantum communication based on a decoherence-free subspace using a counter-propagating weak coherent light pulse}

\author{Hidetoshi Kumagai}
\email{kumagai@qi.mp.es.osaka-u.ac.jp}
\affiliation{Graduate School of Engineering Science, Osaka University,
Toyonaka, Osaka 560-8531, Japan}

\author{Takashi Yamamoto}
\affiliation{Graduate School of Engineering Science, Osaka University,
Toyonaka, Osaka 560-8531, Japan}

\author{Masato Koashi}
\affiliation{Photon Science Center, The University of Tokyo, Bunkyo-ku, Tokyo 113-8565, Japan}

\author{Nobuyuki Imoto}
\affiliation{Graduate School of Engineering Science, Osaka University,
Toyonaka, Osaka 560-8531, Japan}


\begin{abstract}
We study distribution schemes for a polarization entangled photon pair based on a decoherence-free subspace over lossy quantum channels and propose an efficient scheme that is robust against not only collective phase noises but also general collective noises for two qubits. While the proposed scheme employs two photons to build the decoherence-free subspace, the success probability is proportional to the channel transmittance of a single photon with the aid of a counter-propagating weak coherent light pulse. The key ingredient in the scheme is found to be the reciprocity of the channel. The proposed scheme shares the rest of the properties with the previously proposed schemes, i.e., it can be realized by linear optical elements and it is robust against the fluctuations in the optical circuits used by the two parties.


\end{abstract}

\maketitle
\section{\label{sec:level1}Introduction}
Faithful distribution of photonic entangled states among distantly located parties is one of the important issues in the field of quantum information processing, such as quantum key distribution (QKD) \cite{BB84,GisinRMP}, quantum teleportation \cite{teleportation}, and quantum computation \cite{Raussendorf}. Embedding the quantum states into decoherence-free subspace (DFS) consisting of a number of photons or other qubits is considered to be one of the possible schemes to achieve this task \cite{Zanardi, Lidar, Kielpinski, Viola, Ollerenshaw, Mohseni, Roos}. So far, several schemes have been proposed \cite{Walton,Boileau1,TYprl} and experimentally demonstrated \cite{Kwiat,Bourennane,Boileau2,Chen,Prevedel,TYnjp,TYnat} for quantum communication. The photon loss in the quantum channel, however, seriously degrades the transmission rate of quantum states. For example, when the transmittance of one photon through the channel is $T$ and the number of photons used for the DFS is two, the transmission rate of a signal quantum state is proportional to $T^2$. This $T^2$ dependence severely limits the efficiency in long-distance quantum communications. To overcome the inefficiency of the DFS based scheme, an efficient entanglement distribution scheme based on a two-photon DFS, in which the transmission rate is proportional to $T$, has been proposed and experimentally demonstrated \cite{RI}. The key idea in this scheme is the use of a reference photon from a weak coherent  light pulse (WCP) that is counter-propagating through the channel to implement an effective two-photon DFS. However, the protocol is limited for protecting the state only over the collective phase noise in the channel. 

In this paper, we propose a scheme that is an extension of the above two-photon DFS scheme \cite{RI} and show that the present scheme is robust against general collective noises including not only phase shifts but also polarization rotations, while the channel transmission rate of the photon is proportional to $T$. The scheme is realized by linear optical elements and is robust against not only channel noises but also the instability in the optical circuits used by two parties.

This paper is organized as follows. In Sec.\ \ref{opticalfiber}, we discuss the polarization-state transformation for the forward- and the backward-propagating photon in lossy optical media including optical fibers, and introduce a useful relation between these transformations for describing the present scheme. In Sec.\ \ref{entanglementdistribution}, we first introduce the previously proposed protocols based on a two-qubit DFS, and then we provide an extended protocol that is robust against 
the general collective noise channels. In Sec.\ \ref{conclusion}, we give a summary and conclusion. 

\section{Polarization-state transformation in an optical fiber}
\label{opticalfiber}
In this section, we review a relationship between the polarization-state transformations in an optical fiber for forward-propagating (Alice to Bob) and backward-propagating (Bob to Alice) photons. Our derivation is based on an assumption that an optical fiber can be regarded as a sequence of lossy birefringent elements.

\begin{figure}[tbp]
\centering
\includegraphics[width=6cm]{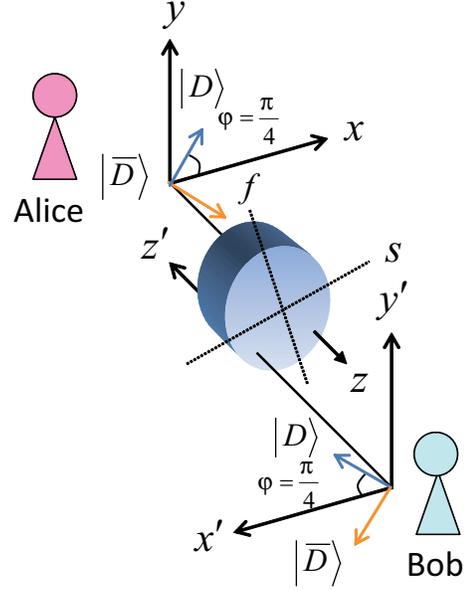}
\caption{(Color online) Coordinate systems for the forward propagation ($xyz$) and the backward propagation ($x^\prime y^\prime z^\prime$). Photons propagate along the $z$ ($z^\prime$) axis. A birefringent element is located with its surface perpendicular to the $z$ ($z^\prime$) axis.}
\label{fig:coordinate}
\end{figure}

First we introduce the coordinate systems used in this paper for describing the polarization states for forward- and backward-propagating photons. We assign two right-handed coordinate systems $xyz$ and $x^\prime y^\prime z^\prime$, which are used for forward- and backward-propagating photons, respectively. As shown in Fig.\ \ref{fig:coordinate}, the forward-propagating photons travel along the $z$ axis and the backward-propagating photons travel along the $z^\prime$ axis. We choose $y$ and $y^\prime$ axes to be in the same direction, and $x$ and $x^\prime$ axes to be in opposite directions. 

A linearly polarized state of a photon with electric field vector along $x$ axis and $y$ axis is represented by $\ket{x}$ and $\ket{y}$, respectively.
The relative phase between $\ket{x}$ and $\ket{y}$ is chosen such that
$\cos\varphi\ket{x}+\sin\varphi\ket{y}$ represents the state linearly polarized in the direction with angle $\varphi$ from $x$ axis. When $\varphi=\pi/4$ and $-\pi/4$, the states are represented by $\ket{D}$ and $\ket{\bar{D}}$, respectively, as shown in Fig.\ \ref{fig:coordinate}.
An arbitrary polarization state can be written as
\begin{align}
\ket{\psi}&=V_{x}\ket{x}+V_{y}\ket{y} \label{eq:1}\\
&=
\begin{pmatrix}
V_{x} \\  V_{y}
\end{pmatrix},
\end{align}
where $V_x$ and $V_y$ are complex numbers satisfying $|V_x|^2+|V_y|^2=1$. The second line shows the corresponding vector representation of the state, which 
coincides with Jones vector \cite{Yariv}. 
We extend the definition of the vector representation to the case where $P(V_x, V_y)\equiv |V_x|^2+|V_y|^2< 1$, such that it represents the state whose density operator is $\rho (V_x, V_y)\equiv (1-P(V_x, V_y))\ket{0}\bra{0}+\ket{\psi}\bra{\psi}$. Here $\ket{0}$ is the vacuum and $\ket{\psi}$ is given by Eq.\ (\ref{eq:1}). 
Then, the effect of any linear passive optical component is represented by a complex matrix $M$ satisfying $M^\dagger M\le \bf{1}$, which transforms the input state $\rho (V_x, V_y)$ to the output state $\rho (V'_x, V'_y)$ as
\begin{align}
\begin{pmatrix}
V'_{x} \\  V'_{y}
\end{pmatrix}
=M\begin{pmatrix}
V_{x} \\  V_{y} 
\end{pmatrix}. 
\end{align}
We call $M$ a transformation matrix here and henceforth. Note that, since $\rho (e^{i\phi}V_x, e^{i\phi}V_y)=\rho (V_x, V_y)$, $M$ and $e^{i\phi}M$ represents the same physical transformation.

\begin{figure}[htbp]
\centering
\includegraphics[width=7.5cm]{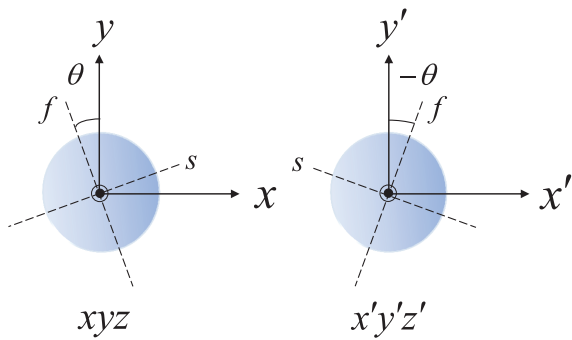}
\caption{(Color online) The inclination of the $f$ axis in $xyz$ and $x^\prime y^\prime z^\prime$ coordinate systems. The angle between the $f$ axis and the $y$ axis is $\theta$ in the $xyz$ coordinate system and $-\theta$ in the $x^\prime y^\prime z^\prime$ coordinate system.}
\label{fig:coordinate2}
\end{figure}

We first introduce a specific example of the transformation matrices for a forward-propagating photon through a birefringent element with a polarization dependent loss. 
The slow and fast axes of the birefringent element are represented by $s$ and $f$, respectively. As shown in Fig.\ \ref{fig:coordinate2}, the angle between the $y$ axis and the $f$ axis is $\theta$, where the angle is measured in a counter-clockwise way; i.e., the positive rotation direction is defined to be from positive $x$ axis to positive $y$ axis. 
The polarization state along $s$ and $f$ axes is represented by $\ket{s}$ and $\ket{f}$, respectively, where 
\begin{equation}
\ket{s}=\cos\theta\ket{x}+\sin\theta\ket{y} 
\end{equation}
and 
\begin{equation}
\ket{f}=-\sin\theta\ket{x}+\cos\theta\ket{y}. 
\end{equation}
We introduce the representation $(V_s,V_f)$ in the basis $\{\ket{s},\ket{f}\}$ from the relation $V_s\ket{s}+V_f\ket{f}=V_x\ket{x}+V_y\ket{y}$, or equivalently,
\begin{align}
\begin{split}
\begin{pmatrix}
V_{s}  \\  V_{f}
\end{pmatrix}
=R(-\theta)
\begin{pmatrix}
V_{x}  \\  V_{y}
\end{pmatrix},
\end{split}
\label{eq:crystallot}
\end{align}
where 
\begin{align}
R(\theta)\equiv \begin{pmatrix}
\cos\theta & -\sin\theta  \\
\sin\theta & \cos\theta
\end{pmatrix}.
\end{align}

After the photon passes through the birefringent element, the state is altered by polarization-dependent phase shifts and photon losses as $\ket{s}\rightarrow \gamma_s e^{-i\phi} \ket{s}$ and $\ket{f}\rightarrow \gamma_f e^{i\phi} \ket{f}$, where $\gamma_s$ and $\gamma_f$ are non-negative real numbers satisfying $\gamma_s\le1$ and $\gamma_f \le 1$. The matrix representation of the state transformation is described as
\begin{equation}
\begin{pmatrix}
V_{s}^\prime  \\  V_{f}^\prime
\end{pmatrix}
=W(\gamma_s, \gamma_f, \phi)
\begin{pmatrix}
V_{s}  \\  V_{f}
\end{pmatrix}, 
\label{eq:phaseshift}
\end{equation}
where 
\begin{equation}
W(\gamma_s, \gamma_f, \phi)\equiv\begin{pmatrix} \gamma_s e^{-i\phi} & 0  \\ 0 & \gamma_f e^{i\phi}\end{pmatrix}.
\end{equation}
The state representation in the basis $\{\ket{x}, \ket{y}\}$ is written as
\begin{equation}
\begin{pmatrix}
V_{x}^\prime  \\  V_{y}^\prime
\end{pmatrix}
=R(\theta)
\begin{pmatrix}
V_{s}^\prime  \\  V_{f}^\prime
\end{pmatrix}.
\end{equation}
Hence the transformation matrix $M$ in the basis $\{\ket{x}, \ket{y}\}$ is given by
\begin{align}
M =
R(\theta)W(\gamma_s, \gamma_f, \phi) R(-\theta).  
\label{eq:Jonesmatrix}
\end{align}

Next we calculate the transformation matrix $\overleftarrow{M}$ for the backward-propagating photon passing through the same medium. 
 In this case, we use the $x^\prime y^\prime z^\prime$ coordinate system. The angle between the $y^\prime$ and $f$ axes is $-\theta$, while $W(\gamma_s, \gamma_f, \phi)$ is unchanged. This leads to 
\begin{align}
\overleftarrow{M}=R(-\theta)W(\gamma_s, \gamma_f, \phi) R(\theta).
\end{align}

Now it is easy to confirm the relations $R(\theta)=Z R^T(\theta)Z$ and $W(\gamma_s, \gamma_f, \phi)=Z W^T(\gamma_s, \gamma_f, \phi)Z$, where $Z$ is a Pauli matrix written as
\begin{equation}
Z=\begin{pmatrix}1 & 0 \\ 0 & -1 \end{pmatrix}. \nonumber
\end{equation} 
Therefore, the relationship between the transformations for forward-propagating and backward-propagating photons is found as 
\begin{equation}
\overleftarrow{M}=ZM^T Z.
\label{eq:forwardandbackward}
\end{equation}
Note that this formula is true for our specific choice of the coordinate systems. In general, the relation (\ref{eq:forwardandbackward}) can be represented by $\overleftarrow{M}=UM^T U^{\dagger}$ with unitary matrix $U$, where $U=R(\alpha)Z$ and $\alpha$ depends on the choice of the coordinate systems \cite{Yoshida,Pistoni}.

The above relation of the single-element transformation between forward and backward propagations is extended to $N$ elements. Suppose that the overall transformation matrix of the sequence of $N$ birefringent elements for a forward-propagating photon is given by $M=M_NM_{N-1}\cdots M_j\cdots M_2M_1$, where $M_j$ stands for the transformation matrix of the $j$th birefringent element. The corresponding transformation matrix for a backward-propagating photon is $\overleftarrow{M}=\overleftarrow{M}_1\overleftarrow{M}_2\cdots\overleftarrow{M}_j \cdots
\overleftarrow{M}_{N-1}\overleftarrow{M}_N$. Because of the relation (\ref{eq:forwardandbackward}), the following equation holds;
 $\overleftarrow{M}=\overleftarrow{M}_1\overleftarrow{M}_2\cdots\overleftarrow{M}_j\cdots
\overleftarrow{M}_{N-1}\overleftarrow{M}_N=Z (M_NM_{N-1}\cdots M_j\cdots M_2M_1)^T Z=ZM^T Z$. This clearly shows that the relation (\ref{eq:forwardandbackward}) is satisfied by a composite system of the birefringent elements. It has been shown that, in general, the relation (\ref{eq:forwardandbackward}) is fulfilled by counter-propagating lights through any reciprocal media \cite{Pistoni}. In Sec. \ref{entanglementdistribution}, we describe our entanglement distribution protocol by using the useful relation (\ref{eq:forwardandbackward}).

\begin{figure}[t]
\centering
\includegraphics[width=6cm]{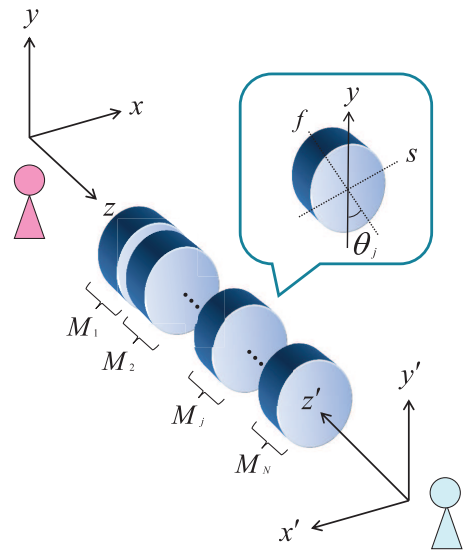}
\caption{(Color online) A sequence of birefringent elements as a model of  the optical fiber. The angle between the $f$ axis and the $y$ axis of the $j$th element is represented by $\theta_j$.}
\label{fig:opticalfiber}
\end{figure}

\section{DFS based quantum communication over collective noise}
\label{entanglementdistribution}

It is well known that logical qubits embedded in a DFS formed by multiple physical qubits are immune to a collective noise. We first introduce a DFS formed by two physical qubits for a collective phase noise channel. Following the convention, hereafter we use the notation of the basis $\{ \ket{H},\ket{V}\}$ instead of $\{\ket{x},\ket{y}\}$, where $\ket{H}$ and $\ket{V}$ represent the horizontally and the vertically polarized single photon state, respectively. A phase-shift channel transforms the states as $\ket{H}\rightarrow e^{-i\phi}\ket{H}$ and $\ket{V}\rightarrow e^{i\phi}\ket{V}$. The corresponding transformation matrix is given by $W(1,1,\phi)$. A photon in the state $\alpha\ket{H}+\beta\ket{V}$ is transformed into  $e^{-i\phi}(\alpha\ket{H}+e^{2i\phi}\beta\ket{V})$ by the phase shift.  If the phase shift $\phi$ varies with time and is unknown, the state is distorted. On the other hand, when the state is encoded into two-photon state $\alpha\ket{HV}+\beta\ket{VH}$ and each photon is considered to be altered by the same phase shift represented by $W(1,1,\phi)$, the state is unchanged as $\alpha\ket{HV}+\beta\ket{VH}$. Thus the logical qubit is protected in the two-qubit DFS spanned by the basis $\{\ket{HV}, \ket{VH} \}$ against the collective phase noise. It is well known that the noise in the optical fiber is mainly caused by the fluctuation of the birefringence, which varies slowly with time. Thus the DFS scheme is useful for quantum communication over optical fibers. 

In this section, we deal with two types of channels; one is a collective phase noise channel and the other is a general collective noise channel. The transformation matrix for each photon through the collective phase noise channel is written as $W(\gamma_s,\gamma_f,\phi)$, which includes phase shift and polarization dependent losses. In the case of the general collective noise channel, the transformation matrix is arbitrary. We assume the relation (\ref{eq:forwardandbackward}) for transformation matrices of those channels. Since the optical fibers are known to be reciprocal media, such an assumption is valid in optical fiber communications. 

In the following subsections, we introduce a protocol that employ one collective phase noise channel and one that employs two general collective noise channels \cite{TYprl,TYnjp}. Then we introduce a protocol using counter-propagating photons, which boosts up the efficiency of the scheme for one collective phase noise channel \cite{RI}. Finally we propose a new scheme that boosts up the efficiency of the scheme for two general collective noise channels. In all examples described in this section, we assume that the fluctuations in the channels are so slow that the transformation matrices do not vary with time.

\subsection{Single-qubit distribution protocol over collective phase noise}
\label{IIIa}

\begin{figure}[tbp]
\centering
\includegraphics[width=7.5cm]{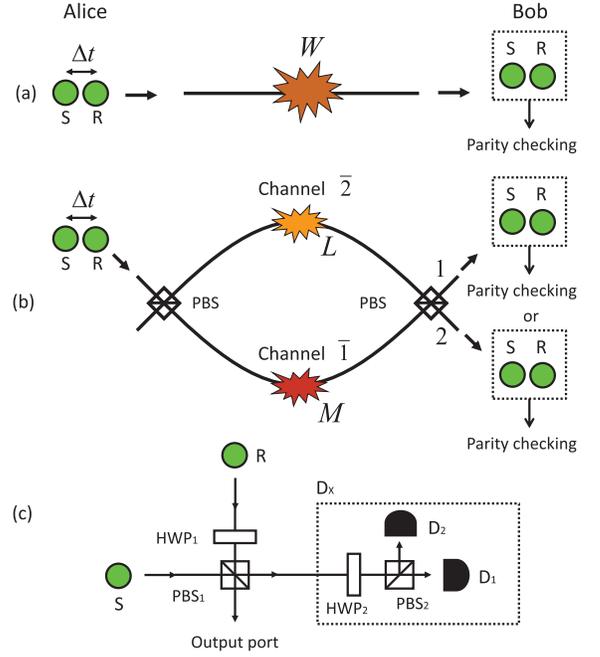}
\caption{(Color online) Schematic diagram of a single qubit distribution using DFS against (a) a collective phase noise channel and (b) two general collective noise channels. Alice is given the signal photon $S$ and the reference photon $R$ in the fixed state $\ket{D}$. The time difference between the signal and the reference photon is $\Delta t$. In the diagram (b), two photons are split into channel $\bar{1}$ and channel $\bar{2}$ by a PBS. The events where two photons appear together in either port 1 or port 2 are selected for the extraction of the state protected by DFS. (c) Linear optical implementation of the parity checking and decoding. Half wave plates $\text{HWP}_1$ and $\text{HWP}_2$ are rotated by $\pi/4$ and $\pi/8$ from the horizontal axis, respectively. The apparatus $\text{D}_X$ surrounded by the dotted box, which includes the photon detector $\text{D}_1$ and $\text{D}_2$, is used for the projection measurement on the basis \{$\ket{D}, \ket{\bar{D}}$\}. The time difference $\Delta t$ between the signal and reference  photons is compensated in advance by using  an optical delay line not shown in the figure.}
\label{fig:DFS1}
\end{figure}

A simple realization of the single-qubit distribution over collective phase noise based on linear optical elements has been proposed in \cite{TYprl}. The procedure of the scheme is as follows: the sender Alice is given a signal photon $S$ in $\alpha\ket{H_S}+\beta\ket{V_S}$ and prepares a reference photon $R$ in a fixed state $\ket{D_R}=\frac{1}{\sqrt{2}}(\ket{H_R}+\ket{V_R})$, where the subscripts inside $\ket{\cdot}$ represent signal and reference. The time difference between the reference photon and the signal photon are separated in time by $\Delta t$ as shown in Fig.\ \ref{fig:DFS1}(a). After the transmission of the two photons through the channel, the state is described as
\begin{align}
\frac{1}{\sqrt 2}[ &\gamma_s\gamma_f(\alpha\ket{V_RH_S}+\beta\ket{H_RV_S}) \notag\\
+&\alpha \gamma^2_s e^{-2i\phi}\ket{H_RH_S}+\beta \gamma^2_f e^{2i\phi}\ket{V_RV_S} ].
\label{eq:two-qubit_DFS1}
\end{align}
The state $\alpha\ket{V_RH_S}+\beta\ket{H_RV_S}$ is in the two-qubit DFS and is invariant under the collective phase noise. 
The projection onto the subspace spanned by the basis $\{\ket{V_RH_S}, \ket{H_RV_S}\}$ and the decoding of its state into the initial signal state 
are performed by linear optical parity checking described in Fig.\ \ref{fig:DFS1}(c).

As shown in Fig. \ref{fig:DFS1}, the linear optical circuits for parity checking aims at the extraction of the state $\alpha\ket{V_RH_S}+\beta\ket{H_RV_S}$ followed by decoding to $\alpha\ket{H_S}+\beta\ket{V_S}$, which works as follows \cite{Pan, TYpra, Pittman}: First one transforms  the polarization of the reference photon $R$ as $\ket{H_R} \rightarrow \ket{V_R}$ and $\ket{V_R} \rightarrow \ket{H_R}$ by $\text{HWP}_1$, followed by sending the photon $R$ to one port of $\text{PBS}_1$ and the photon $S$ to the other port. 
The apparatus $\text{D}_X$, which consists of $\text{HWP}_2$ rotated by $\pi/8$, $\text{PBS}_2$, and photon detectors $\text{D}_1$ and $\text{D}_2$, measures the incoming photons. The photon detection at $\text{D}_1$ and $\text{D}_2$ corresponds to the projection onto the states $\ket{D}$ and $\ket{\bar{D}}$, respectively, when $\text{D}_X$ receives a single photon. In the case where the input state is $\alpha \ket{V_R H_S} +\beta\ket{H_R V_S}$, the state just after the $\text{PBS}_1$ is  $\alpha\ket{H H} +\beta\ket{V V}$. When the photon detection at $\text{D}_1$ and $\text{D}_2$ occur, the state in the output port becomes $\alpha\ket{H} +\beta\ket{V}$ and $\alpha\ket{H} -\beta\ket{V}$, respectively. 
Performing the phase shift $\pi$ only in the case of the photon detection at $\text{D}_2$, we obtain state $\alpha\ket{H}+\beta\ket{V}$, identical to the initial state. On the other hand,  when the input state is in the subspace spanned by $\{\ket{H_R H_S}, \ket{V_R V_S}\}$, two photons leave $\text{PBS}_1$ together from one of the ports, which leads to two- or zero- photon detection in $\text{D}_X$. Thus we can perform the parity checking by linear optics and photon detection. While the photon loss and inefficiency of the detectors leads to the unexpected vacuum in output port, such events can be eliminated by the postselection of the events where the output port is not in the vacuum. 

We should also mention that if one or both of  the input modes include two or more photons, the scheme in Fig.\ \ref{fig:DFS1}(c) results in errors that are  not eliminated by the postselection. In such cases, a single photon detection by the apparatus $\text{D}_X$ may leave a single photon in the output port, but its polarization state will be different from the one intended in the parity checking scheme. In subsection \ref{IIIc} and \ref{IIId}, we discuss protocols using a weak coherent light pulse as the reference, in which case consideration on the rate of such erroneous events  is necessary.

\subsection{Single-qubit distribution protocol over general collective noise}
\label{IIIb}

The scheme in the previous subsection protects a qubit against a collective phase noise channel, but a small modification achieves the protection of the signal state against the general collective noise channels, if we are allowed to use two of such channels as in Fig.\ \ref{fig:DFS1}(b) \cite{TYprl}. The two channels  are combined by a PBS at the sender and by another PBS at the receiver. Let the transformation matrices of the channels $\bar{1}$ and $\bar{2}$ be
\begin{equation}
M=
\begin{pmatrix}
m_1 & m_2 \\ m_3 & m_4
\end{pmatrix}
\text{and}\ 
L=
\begin{pmatrix}
l_1 & l_2 \\ l_3 & l_4
\end{pmatrix}, 
\label{eq:unitaryofthechannel}
\end{equation}
respectively. The signal (reference) photon in state $\ket{H_{S(R)}}$ and $\ket{V_{S(R)}}$ is transformed as $\ket{H_{S(R)}}\rightarrow m_1\ket{H_{S(R)}}+m_3\ket{V_{S(R)}}$ and $\ket{V_{S(R)}}\rightarrow l_2\ket{H_{S(R)}}+l_4\ket{V_{S(R)}}$. After the photons $S$ and $R$ pass through the channels $\bar{1}$ and $\bar{2}$, the state is written as
\begin{widetext}
\begin{align}
\frac{1}{\sqrt 2}[
&\alpha(m_1^2\ket{H_RH_S}_{11}+m_1m_3\ket{H_RV_S}_{12}+m_3m_1\ket{V_RH_S}_{21}+m_3^2\ket{V_RV_S}_{22})\notag \\
+&\beta(m_1l_2\ket{H_RH_S}_{12}+m_1l_4\ket{H_RV_S}_{11}+m_3l_2\ket{V_RH_S}_{22}+m_3l_4\ket{V_RV_S}_{21})\notag \\
+&\alpha(l_2m_1\ket{H_RH_S}_{21}+l_2m_3\ket{H_RV_S}_{22}+l_4m_1\ket{V_RH_S}_{11}+l_4m_3\ket{V_RV_S}_{12})\notag \\
+&\beta(l_2^2\ket{H_RH_S}_{22}+l_2l_4\ket{H_RV_S}_{21}+l_4l_2\ket{V_RH_S}_{12}+l_4^2\ket{V_RV_S}_{11})
].
\end{align}
\end{widetext}
The subscripts outside of $\ket{\cdot}$ represent the output port numbers. When two photons appear in port 1, the state is written as 
$\frac{1}{\sqrt 2}[m_1^2\alpha\ket{H_RH_S}_{11}+m_1l_4(\beta\ket{H_RV_S}_{11}+\alpha\ket{V_RH_S}_{11})+l_4^2\beta\ket{V_RV_S}_{11}]$. 
The state 
$\frac{1}{\sqrt 2}m_1l_4(\beta\ket{H_RV_S}_{11}+\alpha\ket{V_RH_S}_{11})$ 
is invariant under the collective noise. Similarly to the previous scheme, the parity checking shown in Fig.\ \ref{fig:DFS1}(c) achieves the extraction of the signal state $\alpha\ket{H}+\beta\ket{V}$. When the photons $R$ and $S$ appear at port 2, the state is written as 
$\frac{1}{\sqrt 2}[m_3^2\alpha\ket{V_RV_S}_{22}+m_3l_2(\beta\ket{V_RH_S}_{22}+\alpha\ket{H_RV_S}_{22})+l_2^2\beta\ket{H_RH_S}_{22}]$. 
The state 
$\frac{1}{\sqrt 2}m_3l_2(\beta\ket{V_RH_S}_{22}+\alpha\ket{H_RV_S}_{22})$ 
is again invariant under the general collective noise and is decoded into the signal state. As shown in the above discussion, two channels together with PBSs enable us to reject the polarization rotation errors and to extract the signal state. 

The success probability of the case where two photons emerge at port 1 is given by 
$|m_1|^2|l_4|^2/2$, but this value is sensitive to a small change in birefringence of the fiber.
By inserting random unitary operations at both ends of channel $\bar{1}$ and channel $\bar{2}$, we can make the success probability to be a more stable quantity of
$T_1T_2/4$, where $T_1\equiv {\rm Tr}(M^\dagger M)/2$ and $T_2\equiv {\rm Tr}(L^\dagger L)/2$
are the polarization-averaged transmission of the channels. The success probability for two photons leaving port 2 is also given by $T_1T_2/4$, leading to the overall success probability $T_1T_2/2$.

The schemes described in subsections \ref{IIIa} and \ref{IIIb} are able to protect arbitrary unknown states of a qubit, and hence they are also able to protect any correlation that is initially formed between  the input qubit and other systems. Those schemes can thus be used for distributing a maximally entangled state of a qubit pair through channels with collective noises. In the following subsections, we discuss protocols solely intended for such a distribution of  a maximally entangled state, with an added benefit of an improved scaling of the efficiency over the channel transmission.

\subsection{Entanglement distribution protocol over collective phase noise with counter-propagating photons}
\label{IIIc}

A serious drawback of the photonic DFS schemes including the previously introduced schemes is the inefficiency caused by the photon loss in optical fibers. When the transmittance of the channel is $T$ and a two-qubit DFS scheme is performed, the success probability for sending a qubit state is proportional to $T^2$.
In order to overcome the inefficiency, a two-qubit DFS scheme based on a backward-propagating weak coherent light pulse over the collective phase noise channel, whose efficiency is  proportional to $T$, has been proposed and demonstrated in Ref. \cite{RI}. In this subsection, we review the working principle of the scheme in the case where the backward-propagating light is initially a single photon, and then we show that the efficiency is improved by using a coherent light pulse instead of the single photon. 

\begin{figure}[tbp]
\centering
\includegraphics[width=7.5cm]{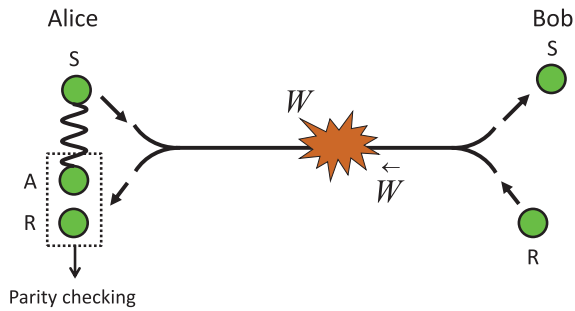}
\caption{(Color online) Schematic diagram of entanglement distribution scheme using the backward-propagating reference photon. Alice prepares the entangled photon pair $\ket{\phi^+_{AS}}$ and sends the photon $S$ to Bob. On the other hand, Bob prepares the reference photon $R$ and sends it to Alice. After receiving the photon $R$, Alice performs parity checking on photons $A$ and $R$ to extract the DFS and decodes back to the entangled state $\ket{\phi^+}$. In order to boost up the efficiency, Bob uses a weak coherent light pulse, in stead of a single photon, as the reference light.}
\label{fig:DFS2-3}
\end{figure}

Suppose that Bob, instead of Alice, sends a reference photon $R$ in the state $\ket{D_R}=\frac{1}{\sqrt 2}(\ket{H_R}+\ket{V_R})$ to Alice as shown in Fig.\ \ref{fig:DFS2-3}. Alice prepares photons $A$ and $S$ in  an entangled state $\ket{\phi^+_{AS}}=\frac{1}{\sqrt 2}(\ket{H_AH_S}+\ket{V_AV_S})$, and sends the signal photon $S$ to Bob through the channel. The transformation matrix $W(\gamma_s, \gamma_f, \phi)$ is used again for the collective phase noise channel. The transformation matrix for the backward propagation is the same as $W(\gamma_s, \gamma_f, \phi)$. After transmission, two photons $A$ and $R$ are at Alice's side and the signal photon $S$ is at Bob's side. Here we use a well-known property that when two qubits $A$ and $S$ are in the entangled state $\ket{\phi^+_{AS}}$, a phase shift on qubit $S$ is equivalent to the same amount of phase shift on qubit $A$.  
Thus the net effect is the same as if photons $A$ and $R$ had passed through the collective phase noise channel as in subsection \ref{IIIa}. 
Thanks to this property, after performing the parity checking on qubit $A$ and $R$ at Alice's side, Alice and Bob obtain the entangled state over the collective phase noise channel.

The efficiency of the protocol using a single photon as the reference photon $R$ is obviously $O(T^2)$. The modification to improve the efficiency is done as follows: Bob sends a coherent light pulse, instead of a single photon, to Alice. 
Let $\mu$ be the average photon number of the coherent pulse received by Alice, after passing through the channel with transmission $T$. The probabilities of one photon and two or more photons are contained in the coherent light pulse at Alice's side are $P_1=O(\mu)$ and $P_m=O(\mu^2)$, respectively. 
In this protocol, the successful events accepted by the linear optical parity checking consists of two cases; (i) one photon is in mode $A$ and one photon is in the reference mode $R$, and (ii) two or more photons are in the reference mode $R$. Since mode $A$ always has a single photon, the probability of the case (i) is $O(\mu)$ and that of the case (ii) is $O(\mu^2)$. As described in subsection \ref{IIIa}, the case (ii) causes the degradation of the fidelity. Thus $O(\mu) \gg O(\mu^2)$ has to hold for high fidelity entanglement distribution, which leads to the condition $\mu\ll 1$. Although the condition $\mu\ll 1$ must be satisfied, $\mu$ can be chosen independent of the channel transmittance $T$. Thus the overall success probability, which is $O(\mu T)$, is proportional to $T$. 
A detailed calculation of the efficiency, the fidelity, and  a trade-off between these values of the scheme with respect to the value of $\mu$ are shown in Appendix.
The advantage of the scheme in the case of low $T$ regime has been experimentally demonstrated in Ref. \cite{RI}.

One might wonder why we cannot apply the same technique to the forward propagation protocol in subsection \ref{IIIa} to improve the efficiency. However, as long as we use the linear optical parity checking in Fig.\   \ref{fig:DFS1}(c) at Bob's side, we do not obtain the efficiency $O(T)$ by using WCP as the reference photon. Suppose that Bob receives a WCP with mean photon number $\mu$. The probability that each of mode $S$ and mode $R$ has exactly one photon, corresponding to the case (i) above,  is $O(\mu T)$ since the signal photon $S$ must have survived the lossy channel. Therefore $O(\mu T) \gg O(\mu^2)$, which leads to $\mu\ll T$, should be satisfied for a high fidelity. This limits the overall success probability to be $O(T^2)$.

\subsection{Entanglement distribution protocol over general collective noise with counter-propagating photons}
\label{IIId}

\begin{figure}[tbp]
\centering
\includegraphics[width=7.5cm]{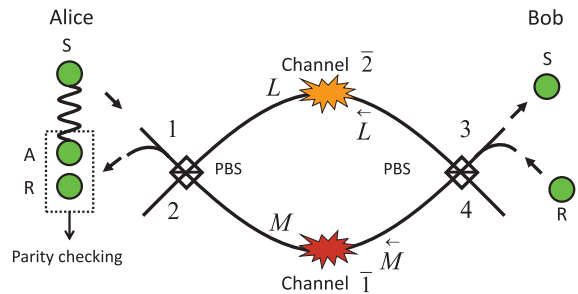}
\caption{(Color online) Entanglement distribution protocol with counter-propagating photons. Only when photon $R$ appears in port 1 and photon $S$ appears in port 3, the state protected by DFS is extracted.}
\label{fig:communicationchannel}
\end{figure}

Here we newly propose an extended scheme that applies the counter propagation protocol to the two-channel scheme introduced in subsection \ref{IIIb} to boost up the efficiency of the two-qubit DFS scheme against the general collective noise channels. The key ingredient in the scheme is the reciprocity relation (\ref{eq:forwardandbackward}) for each channel, which is believed to be satisfied in optical fibers. In the same manner as in the previous subsection, we first describe the working principle when we use a single photon as the reference light.

As shown in Fig.\ \ref{fig:communicationchannel}, Alice prepares the entangled state $\ket{\phi^+_{AS}}$ and sends the signal photon $S$ to Bob. Bob prepares the reference photon $R$ in the state $\ket{D_R}$ and sends it to Alice. The signal photon $S$ is split into two spatial modes by a PBS. The reference photon $R$ is also split into two spatial modes by a PBS. The state just before the signal and reference photons entering channel $\bar{1}$ and $\bar{2}$ is
\begin{align}
\frac{1}{2}(\ket{H_RH_S}_{\bar{1}\bar{1}}&\ket{H_A}+\ket{H_RV_S}_{\bar{1}\bar{2}}\ket{V_A} \notag\\
+&\ket{V_RH_S}_{\bar{2}\bar{1}}\ket{H_A}+\ket{V_RV_S}_{\bar{2}\bar{2}}\ket{V_A}).
\end{align}
Suppose that the transformation matrices of channels $\bar{1}$ and $\bar{2}$ are $M$ and $L$, respectively, in Eq. (\ref{eq:unitaryofthechannel}). The corresponding transformation matrices for backward propagation are $\overleftarrow{M}=ZM^TZ$ and $\overleftarrow{L}=ZL^TZ$ under the reciprocity condition (\ref{eq:forwardandbackward}). 

After the photons $S$ and $R$ pass through the channel $\bar{1}$ and $\bar{2}$, the state is written as 
\begin{align}
\frac{1}{2}
[&(m_1\ket{H_R}_{
\bar{1}}-m_2\ket{V_R}_{\bar{1}})(m_1\ket{H_S}_{\bar{1}}+m_3\ket{V_S}_{\bar{1}})\ket{H_A} \notag  \\
+&(m_1\ket{H_R}_{\bar{1}}-m_2\ket{V_R}_{\bar{1}})(l_2\ket{H_S}_{\bar{2}}+l_4\ket{V_S}_{\bar{2}})\ket{V_A} \notag  \\
+&(-l_3\ket{H_R}_{\bar{2}}+l_4\ket{V_R}_{\bar{2}})(m_1\ket{H_S}_{\bar{1}}+m_3\ket{V_S}_{\bar{1}})\ket{H_A} \notag  \\
+&(-l_3\ket{H_R}_{\bar{2}}+l_4\ket{V_R}_{\bar{2}})(l_2\ket{H_S}_{\bar{2}}+l_4\ket{V_S}_{\bar{2}})\ket{V_A}].
\end{align}
After that, both photons pass through the PBSs. The state then becomes
\begin{widetext}
\begin{align}
\frac{1}{2}[
&(m_1^2\ket{H_R H_S}_{13}+m_1m_3\ket{H_R V_S}_{14}-m_2m_1\ket{V_R H_S}_{23}-m_2m_3\ket{V_R V_S}_{24})\ket{H_A} \notag  \\
+&(m_1l_2\ket{H_R H_S}_{14}+m_1l_4\ket{H_RV_S}_{13}-m_2l_2\ket{V_RH_S}_{24}-m_2l_4\ket{V_R V_S}_{23})\ket{V_A} \notag  \\
+&(-l_3m_1\ket{H_R H_S}_{23}-l_3m_3\ket{H_R V_S}_{24}+l_4m_1\ket{V_R H_S}_{13}+l_4m_3\ket{V_RV_S}_{14})\ket{H_A}  \notag  \\
+&(-l_3l_2\ket{H_RH_S}_{24}-l_3l_4\ket{H_R V_S}_{23}+l_4l_2
\ket{V_R H_S}_{14}+l_4^2\ket{V_R V_S}_{13})\ket{V_A}].
\label{eq:totaloutput}
\end{align}
\end{widetext}

After post-selecting the event where the photons $R$ and $S$ appear at ports 1  and 3, we obtain the state
$\frac{1}{2}[m_1^2\ket{H_R H_S}_{13}\ket{H_A}+m_1l_4(\ket{H_R V_S}_{13}\ket{V_A}+\ket{V_R H_S}_{13}\ket{H_A})+l_4^2\ket{V_R V_S}_{13}\ket{V_A}]$. 
Fortunately, a part of the state 
$\frac{1}{2}m_1l_4(\ket{H_R V_S}_{13}\ket{V_A}+\ket{V_R H_S}_{13}\ket{H_A})$ 
is invariant under the general collective noise. In the same manner as in subsection \ref{IIIb}, Alice can extract the state by using linear optical parity checking in Fig.\ \ref{fig:DFS1}(c). The final state shared between Alice and Bob is the maximally entangled state $\ket{\phi^+}$. As one can see in state (\ref{eq:totaloutput}), unlike the forward propagation protocol shown in subsection \ref{IIIb}, the state of the photons appearing in the other ports, 2 and 4, is not protected over the general collective noise channels. As in subsection \ref{IIIb}, by inserting random unitary operations the overall success probability becomes $T_1T_2/4$, which is half of that in subsection \ref{IIIb} due to the fact that the cases for photons leaving ports 2 and 4 automatically fails. This success probability can then be boosted up by using WCP as the reference
light, from $O(T^2)$ to $O(T)$. The detailed calculations of the efficiency and the fidelity of the scheme are shown in Appendix.  

\begin{figure}[t]
\centering
\includegraphics[width=7.5cm]{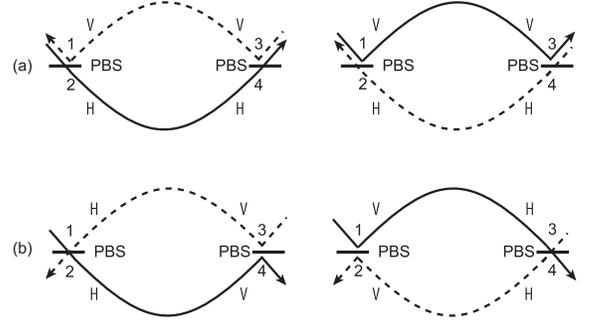}
\caption{Sketch of the trajectories when two photons appear in (a) ports 1 and 3 or (b) ports 2 and 4. The solid arrows and the dotted arrows show the trajectories of the signal and the reference photon, respectively.}
\label{fig:trajectory}
\end{figure}

In the forward-propagating protocol shown in Fig.\ \ref{fig:DFS1}(b), when both photons appear at port 1, there are two possible trajectories: $H\to H /\bar{1}$, implying that an $H$ polarized photon enters channel $\bar{1}$ and leaves in $H$ polarization, and $V\to V /\bar{2}$. If the signal photon has $H$ polarization, it takes the former and the reference photon takes the latter.  If the signal photon has $V$ polarization the trajectories are interchanged. The same argument applies when both photons appear in port 2, with two trajectories $H\to V /\bar{1}$ and $V\to H /\bar{2}$.
In the backward-propagating protocol presented in this subsection, we have four trajectories instead. When the photons appear at ports 1 and 3, those are $H\to H /\bar{1}$ and $V\to V /\bar{2}$ for photon $S$, and $H\gets H /\bar{1}$ and  $V\gets V /\bar{2}$ for photon $R$, where the trajectories for photon $R$ are the time-reversed versions of those for photon $S$. As a result, two possible choices of the trajectories of the two photons, shown in Fig.\ \ref{fig:trajectory}(a), acquire the same phase shift from the channels.
On the other hand, when the photons appear at ports 2 and 4, the relevant trajectories are
$H\to V /\bar{1}$ and $V\to H /\bar{2}$ for photon $S$, and $H\gets V /\bar{2}$ and  $V\gets H /\bar{1}$ for photon $R$, among which no pair is in the time-reversal relation. Hence no state is protected in this case. 

Several remarks are in order for the experimental realization of the proposed scheme using WCP. In practice, a polarization-independent optical circulator with high efficiency is hard to obtain. However, such a device is not required when we use WCP as the reference. In that case, we may replace the optical circulator with small reflectance mirrors, which transmit the signal photon with transmittance close to unity and reflect the reference photons. The low reflectance can simply be compensated by increasing the initial amplitude of the WCP. The optical path length mismatch between the two channels needs to be adjusted within the coherence length of the photons, which is typically far longer than the wavelength of the photons. Similarly to the experiment in Ref. \cite{RI}, the experimental demonstration can be done by using the entangled photon source based on parametric down conversion, linear optical elements and photon detectors. The scheme is also robust against the fluctuations in the optical circuits used for parity checking due to the two-photon interference.

\section{Summary and Conclusion}
\label{conclusion}
We have proposed an efficient entanglement distribution protocol based on a two-qubit DFS, which is robust against not only the collective phase noise channel but also the general collective noise channels. The immunity against the general collective noise is achieved by the rejection of polarization-rotation errors.
This is done by enlarging the Hilbert space via two channels together with PBSs and the postselection. 
The success probability of the scheme is proportional to the channel transmittance of one photon despite the use of two photons for spanning the DFS. The suppression of the photon-loss effect is done by the use of an entangled photon pair and a backward-propagating weak coherent light pulse. 

It is well known that a logical qubit embedded in four-qubit DFS is robust against a general collective noise \cite{Kempe}, which has been demonstrated in Ref. \cite{Bourennane}. If we use such a state to transmit a qubit though optical fiber, the success probability is proportional to $T^4$. On the other hand, our scheme is able to make it proportional to $T$ against the same type of noises. We believe that our robust and efficient quantum communication scheme and the formal study of the polarization state transformation for counter-propagating photons are useful for quantum information processing.

\begin{acknowledgments}
We thank R. Ikuta for helpful discussions. This work was supported by the Funding Program for World-Leading Innovative R $\&$ D on Science and Technology (FIRST), MEXT Grant-in-Aid for Scientific Research on Innovative Areas 21102008, the MEXT Global COE Program, MEXT Grant-in-Aid for Young Scientists (A) 23684035, JSPS Grant-in-Aid for Scientific Research (A) 25247068 and (B) 25286077.
\end{acknowledgments}

\appendix*
\section{}

\begin{figure}[bt]
\centering
\includegraphics[width=7.5cm]{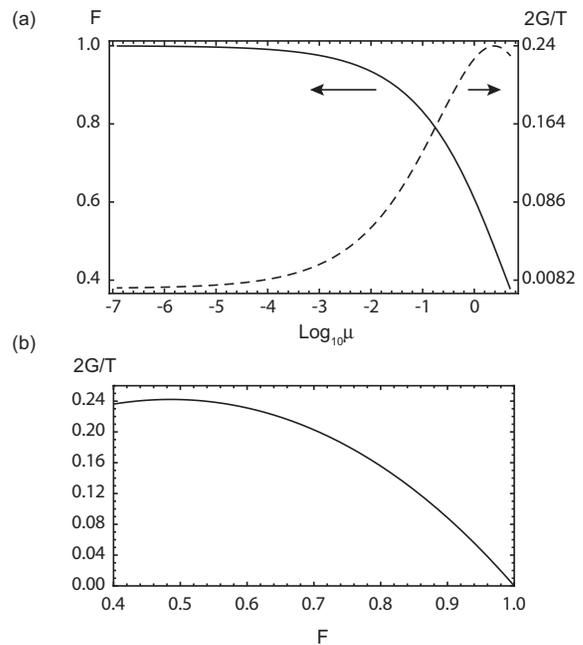}
\caption{Trade-off between the fidelity and the efficiency of the present scheme.}
\label{fig:trade-off}
\end{figure}

In the appendix, we describe the trade-off between the fidelity and the efficiency of the present scheme using counter-propagating weak coherent light pulse. 
Such a trade-off is caused by the coherent-state input of the parity checking in Fig.\ \ref{fig:DFS1}(c). The coherent state is described as $\ket{\alpha}=e^{-|\alpha|^2/2}\sum_n \frac{\alpha^n}{\sqrt{n!}}\ket{n}$, where $\ket{n}$ represents the $n$-photon number state.  The initial entangled state of the photon pair prepared by Alice is described as $\ket{\phi^+}=\frac{1}{\sqrt{2}}(\ket{1}^A_{H}\ket{1}^S_{H}+\ket{1}^A_{V}\ket{1}^S_{V})$, where the superscripts and subscripts of $\ket{\cdot}$ represent the spatial modes and the polarization modes in Fig.\ \ref{fig:DFS1}(c), respectively. 

For simplicity, we assume 
(i) the transmittances $T_1$ and $T_2$ of the channels $\bar{1}$ and $\bar{2}$ do not depend on the polarization state, 
(ii) the random unitary operations are inserted at both ends of channel $\bar{1}$ and channel $\bar{2}$, 
(iii) detectors D$_1$ and D$_2$ have the capability to resolve photon numbers with unit quantum efficiency, and 
(iv) the vacuum component in output port $Y$ and port $3$ in Fig.\ \ref{fig:communicationchannel} is eliminated by a post selection. 
Under these assumptions, multiphoton states in output port $Y$, which are orthogonal to the state $\ket{\phi^+}$, reduces the fidelity of the output state. 

The ancilla state received by Alice is represented by $\ket{m_1\alpha/\sqrt{2}}^R_{H}\ket{l_4\alpha/\sqrt{2}}^R_{V}$ and the entangled state after the distribution is represented by $\ket{\phi^{m_1,l_4}}\equiv\frac{1}{\sqrt{2}}(m_1\ket{1}^A_{H}\ket{1}^S_{H}+l_4\ket{1}^A_{V}\ket{1}^S_{V})$. The unitary transformation performed by HWP$_1$ and PBS$_1$, which we denote $U_1$, transforms the ancilla state $\ket{m_1\alpha/\sqrt{2}}^R_{H}\ket{l_4\alpha/\sqrt{2}}^R_{V}$ to $\ket{l_4\alpha/\sqrt{2}}^Y_{H}\ket{m_1\alpha/\sqrt{2}}^X_{V}$ and the state $\ket{\phi^{m_1,l_4}}$ to $\frac{1}{\sqrt{2}}(m_1\ket{1}^X_{H}\ket{1}^S_{H}+l_4\ket{1}^Y_{V}\ket{1}^S_{V})$. The unitary transformation $U_2$ performed by HWP$_2$ and PBS$_2$ transforms $\ket{m_1\alpha/\sqrt{2}}^X_{V}$ to $\ket{m_1\alpha/2}^{1}_{H}\ket{m_1\alpha/2}^{2}_{V}$ and $\ket{1}^X_{H}$ to $\frac{1}{\sqrt{2}}(\ket{1}^{1}_{H}-\ket{1}^{2}_{V})$. The successful events are the cases where detector $(\text{D}_1, \text{D}_2)$ receives $(\ket{1}^{1}_{H}, \ket{0}^{2}_{V})$ and $(\ket{0}^{1}_{H}, \ket{1}^{2}_{V})$. We only deal with the former case because the fidelity and efficiency of the latter case are equal to the former case. 
The probability of obtaining the former case is calculated to be  
\begin{eqnarray}
&&g(m_1, l_4) 
=||\bra{1}^{1}_H\bra{0}^{2}_V U_2U_1\ket{m_1\alpha/\sqrt{2}}^R_{H}\ket{l_4\alpha/\sqrt{2}}^R_{V}\ket{\phi^{m_1,l_4}}||^2 \nonumber \\
&&-|\bra{0}^Y_V\bra{0}^Y_H\bra{1}^{1}_H\bra{0}^{2}_V U_2U_1\ket{m_1\alpha/\sqrt{2}}^R_{H}\ket{l_4\alpha/\sqrt{2}}^R_{V}\ket{\phi^{m_1,l_4}}|^2
 \nonumber \\
&&=e^{\frac{-|\alpha|^2 |m_1|^2}{2}}\frac{|m_1|^2 }{4}(1+\frac{|\alpha|^2 |l_4|^2}{2}-e^{\frac{-|\alpha|^2 |l_4|^2}{2}})  \nonumber \\
\end{eqnarray}
and the probability of obtaining the state $\ket{\phi^+}$ is calculated to be 
\begin{eqnarray}
 &&h(m_1, l_4) \nonumber \\
 &=&|\bra{\phi^+}\bra{1}^{1}_H\bra{0}^{2}_V U_2U_1\ket{m_1\alpha/\sqrt{2}}^R_{H}\ket{l_4\alpha/\sqrt{2}}^R_{V}\ket{\phi^{m_1,l_4}}|^2 \nonumber \\
&=&e^{\frac{-|\alpha|^2 (|m_1|^2+|l_4|^2)}{2}}\frac{|\alpha|^2|m_1|^2 |l_4|^2}{4}.
\end{eqnarray}

The random unitary operations at both ends of channel $\bar{1}$ and channel $\bar{2}$ achieve uniform probability distribution of $|m_1|^2$ and $|l_4|^2$. Thus the efficiency and the fidelity values are calculated to be 
\begin{eqnarray}
G&=&\int^{2T_1}_{0}\int^{2T_2}_{0}d|m_1|^2 d|l_4|^2 g(m_1, l_4)/4T_1T_2
 \nonumber \\
&=&
\frac{1}{2 |\alpha|^6 T_1 T_2}\left( 1-e^{-|\alpha|^2 T_1}(1+|\alpha|^2 T_1)\right)  \nonumber \\
&& \times
\left(e^{-|\alpha|^2 T_2}-1+|\alpha|^2 T_2 +|\alpha|^4 T^2_2/2 \right)
\end{eqnarray}
and 
\begin{eqnarray}
F&=&\frac{1}{G}\int^{2T_1}_{0}\int^{2T_2}_{0}d|m_1|^2 d|l_4|^2 h(m_1, l_4)/4T_1T_2
 \nonumber \\
&=&
\frac{2 \left(1- e^{-|\alpha|^2 T_2}(1+|\alpha|^2 T_2)\right)}{e^{-|\alpha|^2 T_2}- 1+|\alpha|^2 T_2 +|\alpha|^4 T^2_2/2},
\end{eqnarray}
respectively. Fig.\ \ref{fig:trade-off}(a) shows the fidelity $F$ and the efficiency $2G/T$ as a function of the average photon number $\mu$ of the coherent light in the case where $T_1=T_2=T$ and $\mu=T|\alpha|^2$. This clearly indicates the condition $\mu \ll 1$ to achieve high fidelity. The trade-off between the fidelity and the efficiency is shown in Fig \ \ref{fig:trade-off}(b).

\end{document}